\documentclass[runningheads]{llncs}
\usepackage{graphicx}
\usepackage{graphicx}
\usepackage{booktabs}
\usepackage{multicol}
\usepackage{subfig}

\begin{document}
\title{Brain Tumor Segmentation and Survival Prediction using Automatic Hard mining in 3D CNN Architecture}
%
%
\author{Vikas Kumar Anand\inst{1}\orcidID{0000-0002-7109-1638} \and
Sanjeev Grampurohit\inst{1}\and
Pranav Aurangabadkar\inst{1}\and
Avinash Kori\inst{1}\orcidID{0000-0002-5878-3584} \and
Mahendra Khened\inst{1} \and
Raghavendra S Bhat\inst{2}\and
Ganapathy Krishnamurthi\inst{1}\orcidID{0000-0002-9262-7569}}
\authorrunning{V. K. Anand et al.}
%
\institute{Indian Institute of Technology Madras, Chennai 600036, India\\ \email{gankrish@iitm.ac.in}\\
\and 
Intel Technology India Pvt. Ltd, India
}
\maketitle              
\begin{abstract}
We utilize 3-D fully convolutional neural networks (CNN) to segment gliomas and its constituents from multimodal Magnetic Resonance Images (MRI). The architecture uses dense connectivity patterns to reduce the number of weights and residual connection and is initialized with weights obtained from training this model with BraTS 2018 dataset. Hard mining is done during training to train for the difficult cases of segmentation tasks by increasing the dice similarity coefficient (DSC) threshold to choose the hard cases as epoch increases. On the BraTS2020 validation data (n = 125), this architecture achieved a tumor core, whole tumor, and active tumor dice of  0.744, 0.876, 0.714, respectively. On the test dataset, we get an increment in DSC of tumor core and active tumor by approximately 7\%. In terms of DSC, our network performances on the BraTS 2020 test data are 0.775, 0.815, and 0.85 for enhancing tumor, tumor core, and whole tumor, respectively. Overall survival of a subject is determined using conventional machine learning from rediomics features obtained using generated segmentation mask. Our approach has achieved 0.448 and 0.452 as the accuracy on the validation and test dataset.  
\end{abstract}
\section{Introduction}
A brain tumor is an abnormal mass of tissue that can be malignant or benign. Furthermore, based on risk, a malignant tumor can be classified into two categories, High-Grade Glioma (HGG) and Low-Grade Glioma (LGG). MR imaging is the most commonly used imaging solution to detect the tumor location, size, and morphology. Different modalities of MR imaging enhances separate components of a brain tumor. The Enhancing Tumor (ET) appears as a hyperintense region in the T1Gd image with respect to T1- weighted image and T1Gd image of healthy white matter. Typically resection is performed on the Tumor Core (TC) region. The necrotic region (NCR), non-enhancing region (NET), and ET constitutes the TC.  The NCR and  NET tumor core appear as hypointense areas in T1Gd with respect to T1.
The TC and peritumoral edema (ED) constitutes the WT and describes the disease's full extent. The WT appears as a hyper-intense area in FLAIR. Delineation of the tumor and its component, also called segmentation of tumor region, on several modalities is the first step towards diagnosis. Radiologists carry out this process in a clinical setup, which is time-consuming, and manual segmentation becomes cumbersome with an increase in patients' numbers. Therefore, automated techniques are required to perform segmentation tasks and reduce the radiologist effort. The diffused boundary of the tumor and partial volume effect in the MRI further enhance the challenge in the segmentation of the different regions of the tumor on several MR imaging modalities. 
In recent years, Deep Learning methods, especially Convolutional Neural Networks (CNN), have achieved the state of the art results in the segmentation of different tumor components from a different sequence of MR images \cite{kamnitsas2017efficient,pereira2016brain}. Typically, due to the volumetric nature of medical images, organs are being imaged as 3-D entities, and subsequently, we utilize the nature of 3D CNN based architectures for segmentation task.
\par In this manuscript, we have used patch-based 3D encoder-decoder architecture to segmentation brain tumors from MR volumes. We have also used conditional random field and 3D connected component analysis for post-processing of the segmentation maps.
\section {Related Work}
BraTS 2018 winner, Myronenko et al. \cite{myronenko20183d}, has proposed 3D encoder-decoder architecture with variational autoencoder as a regularization for a large encoder. He has used a non-cuboid patch of a fairly bigger size to train the network with a batch size of 1. Instead of using softmax on several classes or several networks for a different class, all three nested tumor sub-regions are being taken as output after sigmoid. The ensemble of the different networks has given the best result. Isenee et al. \cite{isensee2018no}  has used basic U-Net \cite{ronneberger2015u} with minor modifications. They secure second place in BraTS 2018 challenge by care-full training with data augmentation during training and testing time. For training, a $128^3 $ patch has been used with a batch size of 2. Due to the small batch size instance, normalization has been used. They found an increase in performance using Leaky ReLU instead of the ReLU activation function. BraTS 2019 winner Jiang et al. \cite{jiang2019two} have used two-stage cascaded U-Net architecture to segmentation brain tumors. Segmentation maps obtained in the first stage are being fed to the second stage along with inputs of the first stage. They have also used two decoders in the second stage to get two different segmentation maps. The loss function incorporates all losses that occur due to these segmentations. Data augmentation during training and testing has further improved performance. Data sampling, random patch-size training as a data processing method, semi-supervised learning, architecture development, and fusion of results as a  model devising methods and warming-up learning and multi-task learning optimizing processes have used as different tricks by \cite{zhao2019bag} for 3D brain tumor segmentation. Bag of tricks for segmentation has secured second place in BraTS 2019 challenge.
   \par  This work utilizes a single 3D encoder-decoder architecture for segmentation for different components of a brain tumor. We have used a smaller patch size and hard mining to train our model. Smaller patch size gives us leverage to deploy our model on a smaller GPU, and the hard mining step finds the hard example during training for weighting the loss function. We have not used additional training data and used only data that are provided by challenge organizers.  

\section{Materials and Methods}
A 3D fully convolutional neural network (3DFCNN) \cite{kori2018ensemble} is devised to segment brain tumors and its constituents ET, NER, NET, and ED, from multi-parametric  MR volume. This network is used to achieve semantics segmentation task. Each pixel or volex, which is fed to the network, is assigned with a class label by model. This network has dense connectivity patterns that enhance the flow of information and gradients through the model. This enables us to make a deep network tractable. The predictions produced by the model are smoothened by using Conditional random fields followed by class wise 3D connected component analysis. Post-processing techniques help in decreasing the number of false positives in final segmentation maps.
\subsection{Data}
BraTS 2020 challenge dataset \cite{bakas2017gbm,bakas2017lgg,bakas2017advancing,bakas2018,menze2014multimodal} has been utilized to train the network architecture which is discussed in this manuscript. The training dataset comprises 396 subjects (number of HGG case  = 320 and LGG cases  =  76 ). Each subject has 4 MR sequences, namely FLAIR, T2, T1, T1Gd, and segmentation maps, annotated by an expert on each sequence. Each volume is skull-stripped, rescaled to the same resolution ($1 mm \times  1 mm \times 1 mm$), and co-registered to the common anatomical template.
The BraTS 2020 challenge organizer has issued 125 cases and 166 cases to validate and test the algorithm, respectively. Features such as age, survival days, and resection status are provided separately for the training, validation, and testing phases for 237, 29, and 166 HGG scans.
\subsubsection{Data Pre-processing}
Each volume is normalized to have zero mean and unit standard deviation as a part of pre-processing.
$$img = (img - mean(img))/std (img)$$
$img = $ Only brain region of the volume \\
$mean(img)  = $ mean of a volume\\
$std(img) = $ standard deviation of a volume\\
\subsection{Task1: Brain Tumor Segmentation}
\par \subsubsection{3D fully convolutional encoder-decoder architecture:} 
\begin{figure}
     \centering
     \includegraphics[width = 120mm,height = 110mm]{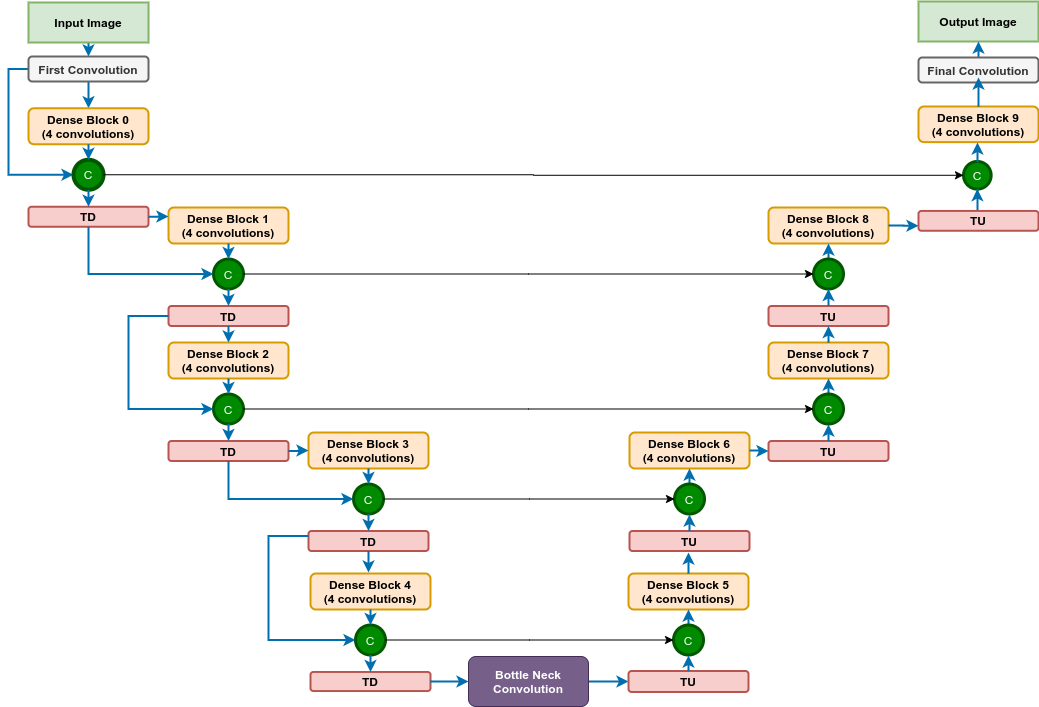}
     \caption{3DFCNN used for segmentation of Brain Tumor and its constituents.  TD: Transition Down block; C: Concatenation block; TU: Transition Up block}
     \label{na}
\end{figure}

The fully convolutional network is used for semantic segmentation task. The input to the network is $64^3$ sized cubes. The network predicts the respective class of voxels in the input cube. Each input to the network has to follow two paths, an encoding, and a decoding path. The encoding architecture of the network comprises Dense blocks along with Transition Down blocks. A series of convolution layers followed by ReLU \cite{nair2010rectified} \& each convolutional layer receives input from all the preceding convolutional layers that make the Dense block. This connectivity pattern leads to the explosion of many feature maps with the network's depth. To overcome the explosion in parameters, set the number of output feature maps per convolutional layer to a small value (k = 4). The spatial dimension of the feature maps is reduced by utilizing the Transition down blocks in the network. The decoding or the up-sampling pathway in the network consists of the
Dense blocks and Transition Up blocks. Transposed convolution layers are utilized to up sample feature maps in the Transition Up blocks. In the decoding section, the dense blocks take features from the encoding part in concatenation with the up-sampled features as input. The network architecture for semantic segmentation is depicted in Figure \ref{na}.
\\
\par \textit{ Patch Extraction}:Each patch is of size $64^3$. These are extracted from the brain.
Many patches are extracted from less frequent classes such as necrosis compared to more frequent classes. This scheme of patch extraction helps in reducing the class imbalance between different classes. The 3DFCNN accepts an input of size $64^3$ and predicts the respective class of the input voxels. There are 77 layers in the network architecture. The effective reuse of the model's features is ensured by utilizing the dense connections among various convolutional layers. The dense connections among layers increase the number of computations, which is subdued by fixing the number of output feature maps per convolutional layer to 4. \\

\par \textit{Training:} The dataset is split into training, validation, and testing in the ratio 70: 20: 10 using stratified sampling based on tumor grade. The network is trained on 205 HGG volumes and 53 LGG volumes. The same ratio of HGG and LGG volumes has been maintained during the validation and testing of the network on held-out data. To further address the issue of class imbalance in the network, the network parameters are trained by minimizing weighted cross-entropy. The weight associated with each class is equivalent to the ratio of the median of the class frequency to the frequency of the class of interest \cite{eigen2015predicting}. The number of samples per batch is set at 4, while the learning rate is initialized to 0.0001 and decayed by a factor of 10\% every-time the validation loss plateaued.\\

\par \textit{Hard Mining:} Our network performed poorly on the hard examples. We have resolved this issue by hard mining such cases and fine-tuned the trained network with these hard mined cases \cite{khened2020generalized}. We implement a threshold-based selection of hard examples. This threshold is obtained using DSC. If a subject has a DSC, which is less than a threshold DSC then this subject is considered a hard example. We choose all such hard examples for a particular set threshold, and we fine-tune our model with these cases. We have chosen two threshold values to fine-tune our model.

\subsection{Task2: Overall Survival prediction}
\begin{figure}
     \centering
     \includegraphics[width = 120mm,height = 30mm]{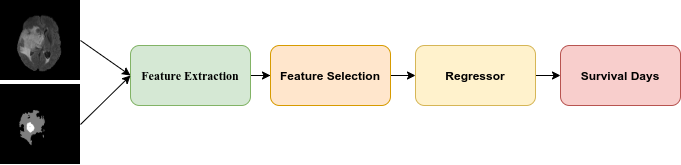}
     \caption{Pipeline used for prediction of overall survival of a patient.}
     \label{surv}
\end{figure}

For training, we have extracted radiomic features using ground truth. There are five types of rediomic features have been extracted for this purpose. These features are first-order radiomic features, which comprises 19 features (mean, median, entropy, etc.); second-order features are Gray Level Co-Occurrence Matrix (GLCM), Gray Level Run Length Matrix (GLRLM),Gray Level Dependence Matrix (GLDM), Gray Level Size Zone Matrix (GLSZM), and Neighboring Gray Tone Difference Matrix (NGTDM), these are altogether 75 different features and 2D and 3D Shape features consists 26 features. We have used pyradicomis \cite{van2017computational} to extract all radiomic features. Using a different combination of segmentation maps, we have extracted 1022 different features. We have assigned an importance value to each feature by using a forest of trees \cite{sklearn_api,scikit-learn}. Thirty-two most important features out of 1022 features are being used to train the Random Forest Regressor (RFR). The pipeline of overall survival prediction is illustrated in Figure \ref{surv}.

\section{Results}
The algorithm is implemented in PyTorch \cite{NEURIPS2019_9015}. The network is trained on NVIDIA GeForce RTX 2080 Ti GPU with Intel Xeon(R) CPU E5-2650 v3 @ 2.30GHz × 20, 32 GB RAM CPU. We have not used any additional data set for training our network. We have taken $64 \times 64 \times 64$ sized non-overlapping patches for training. Four patches from different parametric images such as Flair, T1ce, T2, and T1 images are concatenated as the input of the network.  We have taken the same size of overlapping patches during inference that reduce the edge effect in segmentation maps. 
\par We have reported our network performance on validation and test data sets. The proportion of HGG and LGG in these data sets is not known to us. We have uploaded our segmentation and OS prediction result for validation and test data sets on the BraTS 2020 server. DSC, Sensitivity, Specificity, and Hausdorff Distance have been used as metrics for the network performance evaluation.
\subsection{Segmentation Task:}
\subsubsection{Performance on our network the BraTS 2020 Validation data:}
\begin{figure}[htbp]
     \centering
     \subfloat[][FLAIR image]{\includegraphics[width=.2\linewidth]{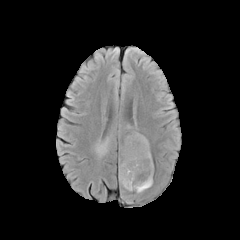}}
     \subfloat[][T1Gd image]{\includegraphics[width=.2\linewidth]{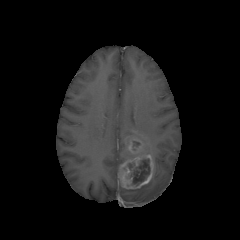}}
     \subfloat[][T1 image]{\includegraphics[width=.2\linewidth]{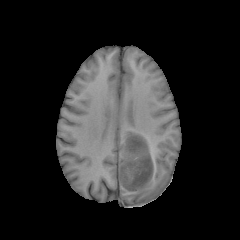}}
     \subfloat[][T2 image]{\includegraphics[width=.2\linewidth]{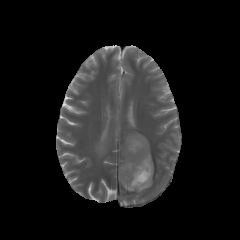}}
     \subfloat[][Segmentation maps]{\includegraphics[width=.211\linewidth]{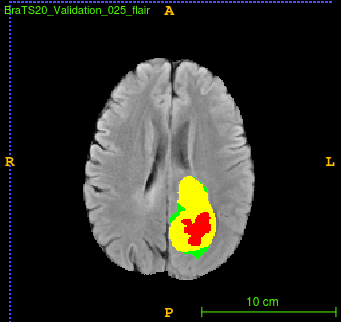}}
     \caption{Segmentation results on validation : Green ,Yellow and Red regions represent edema, enhancing tumor and necrosis respectively.}
     \label{val}
\end{figure}

The model is trained using dice loss to generate segmentation maps. Figure \ref{val} shows different parametric images and the segmentation map obtained on a validation data
\\
On the BraTS validation data (n = 125), the performance of the network is listed in Table \ref{tab:dice_val}.
\begin{table}[htbp]
\caption{Different Metrics for all component of Tumor on the Validation data set}
\label{tab:dice_val}
\resizebox{\textwidth}{!}{%
\begin{tabular}{@{}ccccccccccccc@{}}
\toprule
                      & \multicolumn{3}{c}{DSC}                                                     & \multicolumn{3}{c}{Sensitivity}                                             & \multicolumn{3}{c}{Specificity}                                             & \multicolumn{3}{c}{Hausdroff Distance}                                      \\ \bottomrule
\multicolumn{1}{c|}{} & \multicolumn{1}{c|}{ET} & \multicolumn{1}{c|}{WT} & \multicolumn{1}{c|}{TC} & \multicolumn{1}{c|}{ET} & \multicolumn{1}{c|}{WT} & \multicolumn{1}{c|}{TC} & \multicolumn{1}{c|}{ET} & \multicolumn{1}{c|}{WT} & \multicolumn{1}{c|}{TC} & \multicolumn{1}{c|}{ET} & \multicolumn{1}{c|}{WT} & \multicolumn{1}{c|}{TC} \\\toprule 
Mean                  & 0.71                    & 0.88                    & 0.74                    & 0.74                    & 0.92                    & 0.74                    & 0.99                    & 0.99                    & 0.99                    & 38.31                   & 6.88                    & 32.00                   \\
StdDev                & 0.31                    & 0.13                    & 0.29                    & 0.33                    & 0.14                    & 0.31                    & 0.0005                  & 0.001                   & 0.0005                  & 105.32                  & 12.67                   & 90.55                   \\
Median                & 0.85                    & 0.90                    & 0.89                    & 0.89                    & 0.96                    & 0.89                    & 0.99                    & 0.99                    & 0.99                    & 2.23                    & 3.61                    & 4.24                    \\ \bottomrule
\end{tabular}%
}
\end{table}


\subsubsection{Performance of our network on the BraTS 2020 Test data:}
\begin{figure}[htbp]
     \centering
     \subfloat[][FLAIR image]{\includegraphics[width=.2\linewidth]{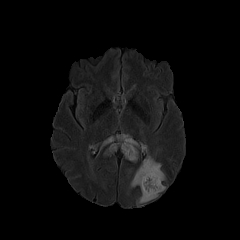}}
     \subfloat[][T1Gd image]{\includegraphics[width=.2\linewidth]{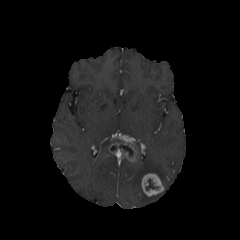}}
     \subfloat[][T1 image]{\includegraphics[width=.2\linewidth]{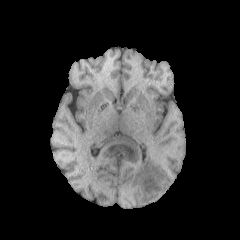}}
     \subfloat[][T2 image]{\includegraphics[width=.2\linewidth]{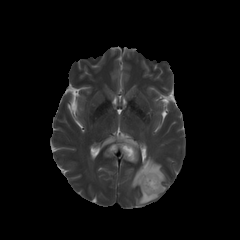}}
     \subfloat[][Segmentation maps]{\includegraphics[width=.211\linewidth]{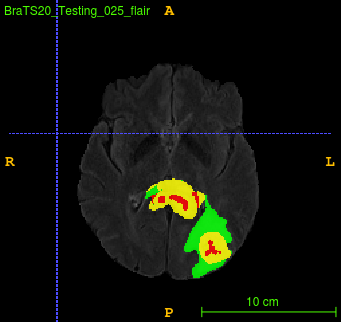}}
     \caption{Segmentation results on test data: Green ,Yellow and Red regions represent edema, enhancing tumor and necrosis respectively.}
     \label{test}
\end{figure}

Figure \ref{test} depicts the different MR images and segmentation maps generated by our network on test data set. Table \ref{tab:test_dice} summarizes all performance metrics obtained on test data set (n = 166). 

\begin{table}[htbp]
\caption{Different Metrics for all component of Tumor on the Test data set}
\label{tab:test_dice}
\resizebox{\textwidth}{!}{%
\begin{tabular}{@{}ccccccccccccc@{}}
\toprule
                      & \multicolumn{3}{c}{DSC}                                                     & \multicolumn{3}{c}{Sensitivity}                                             & \multicolumn{3}{c}{Specificity}                                             & \multicolumn{3}{c}{Hausdroff Distance}                                      \\ \bottomrule
\multicolumn{1}{c|}{} & \multicolumn{1}{c|}{ET} & \multicolumn{1}{c|}{WT} & \multicolumn{1}{c|}{TC} & \multicolumn{1}{c|}{ET} & \multicolumn{1}{c|}{WT} & \multicolumn{1}{c|}{TC} & \multicolumn{1}{c|}{ET} & \multicolumn{1}{c|}{WT} & \multicolumn{1}{c|}{TC} & \multicolumn{1}{c|}{ET} & \multicolumn{1}{c|}{WT} & \multicolumn{1}{c|}{TC} \\\toprule 
Mean                  & 0.776                    & 0.8507                    & 0.815                    & 0.833                    & 0.923                    & 0.838                    & 0.999                    & 0.999                    & 0.999                    & 19.11                   & 8.07                    & 21.276               \\
StdDev                & 0.223                    & 0.164                    & 0.255                    & 0.249                    & 0.153                    & 0.257                    & 0.0004                  & 0.0016                   & 0.0007                  & 74.87                  & 12.21                   & 74.66                   \\
Median                & 0.836                    & 0.902                    & 0.907                    & 0.926                    & 0.968                    & 0.940                    & 0.999                    & 0.999                    & 0.999                   & 2                  & 4.12                    & 3                    \\ \bottomrule
\end{tabular}%
}
\end{table}


\subsection{OS prediction task:} Figure \ref{surv} explains the overall flowchart of the experiment used to find the survival day. Feature extractor module is used to drive all rediomics features using all 4 type sequenced images and corresponding ground truth with several combinations. Before obtaining the importance of a feature for this task, we have standardized the training and validation feature matrix. Feature importance is obtained using a forest of trees. RFR has been used as a regressor for this task. Table \ref{tab:os_val} comprises the different metrics obtained during training and validation.     
\subsubsection{Prediction of OS using training and validation data:}
During validation phase, clinical information of 29 cases are provided to find the OS of case. Table \ref{tab:os_val} contains all performance metrics for OS prediction.  \\
\begin{table}[!htbp]
\caption{Different metrics obtained on training and validation data to evaluate the survival of a patient.}
\label{tab:os_val}
\resizebox{\textwidth}{!}{%
\begin{tabular}{@{}cccccc@{}}
\toprule
           & Accuracy & MSE        & medianSE  & stdSE      & SpearmanR \\ \midrule
Training   & 0.59     & 44611.096  & 13704.672 & 109337.932 & 0.775     \\
Validation & 0.448    & 110677.443 & 22874.178 & 142423.687 & 0.169     \\ \bottomrule
\end{tabular}%
}
\end{table}\\

\subsubsection{Prediction of over all survival using test data :}
For testing phase, clinical information is available for 166 cases. Table \ref{tab:os_test} is the summary of performance metrics of our algorithms for survival prediction.\\
\begin{table}[htbp]
\caption{Different metrics obtained on test data to evaluate the survival of a patient.}
\label{tab:os_test}
\resizebox{\textwidth}{!}{%
\begin{tabular}{@{}cccccc@{}}
\toprule
           & Accuracy & MSE        & medianSE  & stdSE      & SpearmanR \\ \midrule
Test & 0.452    & 4122630758 & 67136.258 & 1142775.390 & -0.014     \\ \bottomrule
\end{tabular}%
}

\end{table}
\section {Conclusion}
This manuscript deals with 2 out of 3 problems posed by the challenge organizer. We have illustrated the use of a fully convolutional neural network for brain tumor segmentation. Overall survival prediction is calculated using generated segmentation maps with conventional machine learning algorithms. We have used the single network from our previous participation in the BraTS 2018 challenge, described in \cite{kori2018ensemble}. This year, we have introduced the hard mining steps during training the network. Our model has achieved DSC of 0.71 and 0.77 on enhancing tumors, 0.88 and 0.85 on the whole tumor, and 0.74 and 0.81 on the tumor core for validation and test data. We have observed that the hard mining step improves DSC for tumor core by 9\%, DSC for ET by 8\%, and the whole tumor by 2\%. Hard mining step makes it easy to learn hard examples by the network during training.
\newpage

%
%
%
\bibliographystyle{unsrt}
\bibliography{main}

\begin{thebibliography}{10}

\bibitem{kamnitsas2017efficient}
Konstantinos Kamnitsas, Christian Ledig, Virginia~FJ Newcombe, Joanna~P
  Simpson, Andrew~D Kane, David~K Menon, Daniel Rueckert, and Ben Glocker.
\newblock Efficient multi-scale 3d cnn with fully connected crf for accurate
  brain lesion segmentation.
\newblock {\em Medical image analysis}, 36:61--78, 2017.

\bibitem{pereira2016brain}
S{\'e}rgio Pereira, Adriano Pinto, Victor Alves, and Carlos~A Silva.
\newblock Brain tumor segmentation using convolutional neural networks in mri
  images.
\newblock {\em IEEE transactions on medical imaging}, 35(5):1240--1251, 2016.

\bibitem{myronenko20183d}
Andriy Myronenko.
\newblock 3d mri brain tumor segmentation using autoencoder regularization.
\newblock In {\em International MICCAI Brainlesion Workshop}, pages 311--320.
  Springer, 2018.

\bibitem{isensee2018no}
Fabian Isensee, Philipp Kickingereder, Wolfgang Wick, Martin Bendszus, and
  Klaus~H Maier-Hein.
\newblock No new-net.
\newblock In {\em International MICCAI Brainlesion Workshop}, pages 234--244.
  Springer, 2018.

\bibitem{ronneberger2015u}
Olaf Ronneberger, Philipp Fischer, and Thomas Brox.
\newblock U-net: Convolutional networks for biomedical image segmentation.
\newblock In {\em International Conference on Medical image computing and
  computer-assisted intervention}, pages 234--241. Springer, 2015.

\bibitem{jiang2019two}
Zeyu Jiang, Changxing Ding, Minfeng Liu, and Dacheng Tao.
\newblock Two-stage cascaded u-net: 1st place solution to brats challenge 2019
  segmentation task.
\newblock In {\em International MICCAI Brainlesion Workshop}, pages 231--241.
  Springer, 2019.

\bibitem{zhao2019bag}
Yuan-Xing Zhao, Yan-Ming Zhang, and Cheng-Lin Liu.
\newblock Bag of tricks for 3d mri brain tumor segmentation.
\newblock In {\em International MICCAI Brainlesion Workshop}, pages 210--220.
  Springer, 2019.

\bibitem{kori2018ensemble}
Avinash Kori, Mehul Soni, B~Pranjal, Mahendra Khened, Varghese Alex, and
  Ganapathy Krishnamurthi.
\newblock Ensemble of fully convolutional neural network for brain tumor
  segmentation from magnetic resonance images.
\newblock In {\em International MICCAI Brainlesion Workshop}, pages 485--496.
  Springer, 2018.

\bibitem{bakas2017gbm}
Spyridon Bakas, Hamed Akbari, Aristeidis Sotiras, Michel Bilello, Martin
  Rozycki, Justin Kirby, John Freymann, Keyvan Farahani, and Christos
  Davatzikos.
\newblock Segmentation labels and radiomic features for the pre-operative scans
  of the tcga-gbm collection. the cancer imaging archive.
\newblock {\em Nat Sci Data}, 4:170117, 2017.

\bibitem{bakas2017lgg}
Spyridon Bakas, Hamed Akbari, Aristeidis Sotiras, Michel Bilello, Martin
  Rozycki, Justin Kirby, John Freymann, Keyvan Farahani, and Christos
  Davatzikos.
\newblock Segmentation labels and radiomic features for the pre-operative scans
  of the tcga-lgg collection.
\newblock {\em The cancer imaging archive}, 286, 2017.

\bibitem{bakas2017advancing}
Spyridon Bakas, Hamed Akbari, Aristeidis Sotiras, Michel Bilello, Martin
  Rozycki, Justin~S Kirby, John~B Freymann, Keyvan Farahani, and Christos
  Davatzikos.
\newblock Advancing the cancer genome atlas glioma mri collections with expert
  segmentation labels and radiomic features.
\newblock {\em Scientific data}, 4:170117, 2017.

\bibitem{bakas2018}
Spyridon Bakas, Mauricio Reyes, Andras Jakab, Stefan Bauer, Markus Rempfler,
  Alessandro Crimi, Russell~Takeshi Shinohara, Christoph Berger, Sung~Min Ha,
  Martin Rozycki, et~al.
\newblock Identifying the best machine learning algorithms for brain tumor
  segmentation, progression assessment, and overall survival prediction in the
  brats challenge.
\newblock {\em arXiv preprint arXiv:1811.02629}, 2018.

\bibitem{menze2014multimodal}
Bjoern~H Menze, Andras Jakab, Stefan Bauer, Jayashree Kalpathy-Cramer, Keyvan
  Farahani, Justin Kirby, Yuliya Burren, Nicole Porz, Johannes Slotboom, Roland
  Wiest, et~al.
\newblock The multimodal brain tumor image segmentation benchmark (brats).
\newblock {\em IEEE transactions on medical imaging}, 34(10):1993--2024, 2014.

\bibitem{nair2010rectified}
Vinod Nair and Geoffrey~E Hinton.
\newblock Rectified linear units improve restricted boltzmann machines.
\newblock In {\em ICML}, 2010.

\bibitem{eigen2015predicting}
David Eigen and Rob Fergus.
\newblock Predicting depth, surface normals and semantic labels with a common
  multi-scale convolutional architecture.
\newblock In {\em Proceedings of the IEEE international conference on computer
  vision}, pages 2650--2658, 2015.

\bibitem{khened2020generalized}
Mahendra Khened, Avinash Kori, Haran Rajkumar, Balaji Srinivasan, and Ganapathy
  Krishnamurthi.
\newblock A generalized deep learning framework for whole-slide image
  segmentation and analysis.
\newblock {\em arXiv preprint arXiv:2001.00258}, 2020.

\bibitem{van2017computational}
Joost~JM Van~Griethuysen, Andriy Fedorov, Chintan Parmar, Ahmed Hosny, Nicole
  Aucoin, Vivek Narayan, Regina~GH Beets-Tan, Jean-Christophe Fillion-Robin,
  Steve Pieper, and Hugo~JWL Aerts.
\newblock Computational radiomics system to decode the radiographic phenotype.
\newblock {\em Cancer research}, 77(21):e104--e107, 2017.

\bibitem{sklearn_api}
Lars Buitinck, Gilles Louppe, Mathieu Blondel, Fabian Pedregosa, Andreas
  Mueller, Olivier Grisel, Vlad Niculae, Peter Prettenhofer, Alexandre
  Gramfort, Jaques Grobler, Robert Layton, Jake VanderPlas, Arnaud Joly, Brian
  Holt, and Ga{\"{e}}l Varoquaux.
\newblock {API} design for machine learning software: experiences from the
  scikit-learn project.
\newblock In {\em ECML PKDD Workshop: Languages for Data Mining and Machine
  Learning}, pages 108--122, 2013.

\bibitem{scikit-learn}
F.~Pedregosa, G.~Varoquaux, A.~Gramfort, V.~Michel, B.~Thirion, O.~Grisel,
  M.~Blondel, P.~Prettenhofer, R.~Weiss, V.~Dubourg, J.~Vanderplas, A.~Passos,
  D.~Cournapeau, M.~Brucher, M.~Perrot, and E.~Duchesnay.
\newblock Scikit-learn: Machine learning in {P}ython.
\newblock {\em Journal of Machine Learning Research}, 12:2825--2830, 2011.

\bibitem{NEURIPS2019_9015}
Adam Paszke, Sam Gross, Francisco Massa, Adam Lerer, James Bradbury, Gregory
  Chanan, Trevor Killeen, Zeming Lin, Natalia Gimelshein, Luca Antiga, Alban
  Desmaison, Andreas Kopf, Edward Yang, Zachary DeVito, Martin Raison, Alykhan
  Tejani, Sasank Chilamkurthy, Benoit Steiner, Lu~Fang, Junjie Bai, and Soumith
  Chintala.
\newblock Pytorch: An imperative style, high-performance deep learning library.
\newblock In H.~Wallach, H.~Larochelle, A.~Beygelzimer, F.~d\textquotesingle
  Alch\'{e}-Buc, E.~Fox, and R.~Garnett, editors, {\em Advances in Neural
  Information Processing Systems 32}, pages 8024--8035. Curran Associates,
  Inc., 2019.

\end{thebibliography}

\end{document}